\documentclass[aps,twocolumn,prd,preprintnumbers, superscriptaddress,floatfix, showpacs,nofootinbib]{revtex4-1} 
\usepackage{amsmath}
\usepackage{amssymb}
\usepackage{hyperref}
\usepackage{graphicx}
\usepackage{xcolor}
\usepackage{subfig}
\usepackage{appendix}
\usepackage[normalem]{ulem}

\begin{document}
\title{\bf Domain Walls From Confining Bubbles:\\ SU(N) Yang-Mills at Finite $\theta$}
\author{Bruno Missoni}\email[Electronic address: ]{bmissoni@sissa.it}
\affiliation{SISSA, International School for Advanced Studies, Via Bonomea 265, 34136, Trieste, Italy}
\affiliation{INFN Sezione di Trieste, Via Bonomea 265, 34136, Trieste, Italy}
\affiliation{IFPU, Institute for Fundamental Physics of the Universe, Via Beirut 2, 34014 Trieste, Italy}
\author{Enrico Morgante}\email[Electronic address: ]{enrico.morgante@units.it}
\affiliation{Dipartimento di Fisica, Università di Trieste, Strada Costiera 11, I-34151 Trieste, Italy}
\affiliation{INFN, Sezione di Trieste, Via Valerio 2, 34127 Trieste, Italy}
\author{Nicklas Ramberg}\email[Electronic address: ]{nramberg@sissa.it}
\affiliation{SISSA, International School for Advanced Studies, Via Bonomea 265, 34136, Trieste, Italy}
\affiliation{INFN Sezione di Trieste, Via Bonomea 265, 34136, Trieste, Italy}
\affiliation{IFPU, Institute for Fundamental Physics of the Universe, Via Beirut 2, 34014 Trieste, Italy}
\date{\today}

\newcommand{\EM}[1]{\textcolor{orange}{EM: #1}}

\begin{abstract}
We study the confinement phase transition in SU($N_{c}$) pure Yang-Mills
theory at finite $\theta \neq 0$ using the Improved Holographic QCD (IHQCD) model.
We show that the critical temperature, as a function of $\theta$ for large but fixed
$N_{c}$, is reduced, thus decreasing the
amount of supercooling in the confinement phase transition. 
Upon completion
of the confinement phase transition, a network of domain walls can be produced,
owing to the multi-branched vacuum structure of Yang-Mills theory at finite
$\theta$. We highlight the potential interplay between the produced domain
walls and the confinement phase transition dynamics. We emphasize that DW production from bubble coalescence in strongly coupled non-conformal FOPTs is a dynamical process of vacuum assignment, hydrodynamics, and local reheating effects, all potentially affecting the approach towards the scaling regime.  Lastly, we demonstrate the level of tuning necessary for potentially interesting imprints from gravitational waves through domain wall annihilation and its interplay with the confinement PT.
\end{abstract}
\preprint{SISSA 12/2026/FISI}
\maketitle

\section{Introduction}

Spontaneous symmetry breaking is expected to happen in the early universe, as fundamental symmetries are restored at high temperature. As the universe cools down, the breaking may manifest itself as a First or Second-Order Phase Transition (FOPT/SOPT), leaving behind a population of topological defects such as monopoles~\cite{tHooft:1974kcl, Polyakov:1974ek, Prasad:1975kr},  cosmic strings~\cite{Vilenkin:1981kz, Vilenkin:1984ib, Kibble:1976sj}, or domain walls~\cite{Roshan:2026xpf, Wei:2022poh}. Crucially, FOPTs in the early universe generate anisotropic stress in the energy-momentum tensor, thereby sourcing Gravitational Waves (GWs) that offer observational access to physics beyond the Standard Model~\cite{Yagi:2011wg, LISA:2017pwj, Janssen:2014dka, Kierkla:2022odc, Kierkla:2023von, Schmitt:2024pby, Sagunski:2023ynd, Lewicki:2024xan, Madge:2023dxc, Biondini:2026uds, Croon:2020cgk, Christiansen:2025xhv, Breitbach:2018ddu, Madge:2018gfl, Barni:2025gnm, Blasi:2023rqi, Baratella:2018pxi, Bringmann:2023opz, Bringmann:2026xcx, Balan:2025uke}.
While the Electroweak and the QCD phase transitions are known to be smooth cross-overs~\cite{Kajantie:1996mn, Bhattacharya:2014ara}, strongly coupled dark sectors may naturally exhibit FOPTs during confinement or chiral symmetry breaking.
The GW signal from a $\mathrm{SU}(N_c)$ Yang-Mills theory~\cite{Agrawal:2025xul, Morgante:2022zvc, Huber:2025qbl, Pasechnik:2023hwv, Reichert:2021cvs, Huang:2020crf, Houtz:2025ogg, Croon:2019iuh} is generally expected to be weak, due to the small amount of supercooling of the metastable deconfined phase. However, because complete non-perturbative end-to-end computations of these FOPTs and their GWs are still lacking, this prediction is not conclusive yet. Alternatively, one can envision strongly coupled QFTs that achieve substantially greater supercooling and, therefore, stronger GW signals in models with conformal, walking-type dynamics ~\cite{Agrawal:2025wvf, Azatov:2020nbe}.

It is natural to extend previous studies of confining PTs to the case of non-zero $\theta$-angle, to assess its quantitative effects on the PT phase diagram. Because the confinement PT is inherently strongly coupled, we employ holographic techniques in this work. In particular, we use the Improved Holographic QCD (IHQCD) model~\cite{Gursoy:2007cb, Gursoy:2007er, Gursoy:2008za, Gursoy:2009jd}  with an axion to describe $SU(N_c)$ YM at finite $\theta$. More recent studies of holographic models and setups to investigate the effects of $\theta$ have been carried out in \cite{Mishra:2026lvq,Bigazzi:2019eks,Bigazzi:2020phm,Csaki:2026arv,Csaki:2026qjl,Bartolini:2016dbk}. Other approaches for studying $SU(N_c)$ YM at finite $\theta$ and its PT structure have been considered in ~\cite{DElia:2012pvq, Bonanno:2023hhp} using lattice QCD at imaginary $\theta$ angle or looking at different topologies ~\cite{Poppitz:2021cxe}.

In this work, we focus on the confinement PTs of $\mathrm{SU}(N_c)$ YM theories at $\theta \neq 0$, after inflation. The dark sector couples to the SM only gravitationally.
In such a setup, we show that the amount of supercooling is reduced as $\theta$ is increased for fixed $N_c$. However, due to the multibranched vacuum structure of YM theory at finite $\theta$, a domain wall (DW) network can be formed upon the completion of the confining PT.
We further argue that DW formation in strongly coupled non-conformal FOPTs involves a hierarchy of hydrodynamic and cooling timescales, going beyond the standard percolation picture.
Finally, we elaborate on the interplay between the DWs and FOPT dynamics and discuss the imprints of $\theta\neq 0$ on the GW spectrum of annihilating DWs ~\cite{Notari:2025kqq, Babichev:2025stm, Blasi:2025tmn, Blasi:2022ayo, Blasi:2023sej, ZambujalFerreira:2021cte}.

This paper is organized as follows: in Sec.~\ref{Section: Review of IHQCD}, we review the IHQCD model at finite $\theta$, both at zero and finite temperature. In Sec.~\ref{Section: Confining PT at finite theta} we discuss the confinement FOPT with finite $\theta$ and $N_{c}$. Furthermore, in Sec.~\ref{Section: DW production between different branches} we explain how the DW network is produced after the confining PT is completed. We consider the GW signal from annihilating DWs in Sec. \ref{Section: GW from annihilating domain walls}. In Sec.~\ref{Section: Discussion and Conclusions} we discuss and conclude our results.

\section{Review of Improved Holographic QCD with finite $\theta$}\label{Section: Review of IHQCD}

Holographic models arise as solutions of ten-dimensional string theory, and they have been successful in the description of confinement, IR dynamics, and thermal phase transitions of strongly-coupled gauge theories (e.g \cite{Witten:1998zw, Alvarez-Gaume:2005dvb, Klebanov:1999tb, Klebanov:2000hb, Klebanov:2000nc, Mateos:2006nu, Horowitz:2007fe, Dillon:2017ctw}). However, when applied to QCD, such top-down constructions typically contain extra Kaluza-Klein (KK) modes, not present in QCD. To avoid extra KK modes, we should consider string theories living in five dimensions, i.e. non-critical string theories. Meanwhile, still inspired by non-critical string theories, here we adopt a more phenomenological approach from \cite{Gursoy:2007cb,Gursoy:2007er,Gursoy:2008za, Gursoy:2009jd}, known as Improved Holographic QCD (IHQCD). In this model, the effective gravitational theory is constructed bottom-up to reproduce general properties of a QCD-like theory, such as confinement, the glueball and meson spectrum, the finite-temperature phase diagram, etc. In this work, we focus on the gravity dual describing a pure $SU(N_c)$ Yang-Mills theory.

The spectrum of the dual gravity theory consists of:
\begin{itemize}
    \item Metric $g_{\mu\nu}$: dual to the energy-momentum tensor $T_{\mu\nu}$.
    \item Dilaton $\Phi$: dual to $\text{Tr}\ F^2$. Asymptotically, it is related to the 't Hooft coupling of the Yang-Mills theory $\lambda_{YM}$.
    \item Axion $a$: dual to $\text{Tr} \ F\wedge F$. Its asymptotic value is equal to the UV $\theta$-angle, i.e. $\theta_{UV}$.
\end{itemize}
The action, written in the Einstein frame, is given by
\begin{equation}
\begin{split}
    S=&-M_{Pl}^3N_c^2\times\\
    &\int_{\mathcal{M}} d^5x \sqrt{g}\bigg[R-\frac{4}{3}(\partial\Phi)^2-\frac{Z(\Phi)}{2N_c^2}(\partial a)^2+V(\Phi)\bigg]\\
&+2M_{Pl}^3N_c^2\int_{\partial\mathcal{M}}d^4x \sqrt{h}\mathcal{K},
\end{split}
\end{equation}
where $M_{Pl}$ is the Planck mass, $N_c$ is the number of colours and $R$ is the 5D Ricci scalar. The boundary term is the famous GHY boundary term \cite{Gibbons:1976ue} and $\mathcal{K}$ is the extrinsic curvature
\begin{equation}
    K_{\mu\nu}=\nabla_{\mu}n_\nu=\frac{1}{2}n^\rho\partial_\rho h_{\mu\nu}\quad \mathcal{K}=h^{ab}K_{ab},
\end{equation}
where $h$ represents the induced metric on the boundary $\partial \mathcal{M}$.
The effective 5D Newton constant is given by $G_5 = 1/(16\pi M_{Pl}^3 N_c^2)$.  

The dilaton potential is determined from the requirement of having a confining background at zero temperature \cite{Gursoy:2007cb,Gursoy:2007er}. We will usually adopt the following potential ($\lambda=e^\Phi$) \cite{Gursoy:2009jd,Morgante:2022zvc} 
\begin{equation}\label{Choice of the potential V(phi)}
    V(\lambda)=\frac{12}{\ell^2}(1+V_0\lambda+V_1\lambda^{4/3}[\log(1+V_2\lambda^{4/3}+V_3\lambda^2)]^{1/2}),
\end{equation}
where $\ell$ is the AdS length and
\begin{equation}
    V_0=\frac{8}{9}b_0, \quad V_2=b_0^4\left(\frac{23+36\frac{b_1}{b_0^2}}{81V_1}\right)^2
\end{equation}
where $b_0=22/(3(4\pi)^2)$ and $b_1/b_0^2=51/121$. This choice comes from setting $\lambda$ proportional to the 't Hooft coupling in the UV \cite{Morgante:2022zvc}. Then $b_0$ and $b_1$ are the coefficients of the asymptotic $\beta$-function
\begin{equation}
    \beta(\lambda)\approx-b_0\lambda-b_1\lambda^2+\mathcal{O}(\lambda^3), \quad \lambda\rightarrow 0\,,
\end{equation} 
up to two loop order. 
Note that, as $\lambda\rightarrow0$, $V(\lambda)\rightarrow12/\ell^2$ and the solutions to Einstein equations will be asymptotically AdS. In the case of pure YM, the $\beta$-function coefficients up to two-loop order are independent of $N_{c}\,,$ see Appendix A in ~\cite{Gursoy:2007cb,Gursoy:2007er}. 
The coefficients $V_1$ and $V_3$ are determined by comparison with the lattice data\footnote{In particular, by matching the glueball spectrum at zero temperature and the pressure and trace anomaly at some reference temperature.} \cite{Gursoy:2009jd}
\begin{equation}
    V_1=14, \quad V_3=170.
\end{equation}
Similarly, the function $Z(\lambda)$ is determined from the lattice data and takes the form
\begin{equation}
    Z(\lambda)=Z_0(1+c_a\lambda^4),
\end{equation}
where $c_a=0.26$ and $Z_0=33.25$ in order to match the pseudoscalar glueball spectrum and the topological susceptibility \cite{Gursoy:2012bt,Gursoy:2009jd}. Note that the above matching was obtained for $N_c=3$. However, we will keep $N_c$ arbitrary in our expressions.
\begin{figure}[t!]
    \centering
    \includegraphics[width=0.85\linewidth]{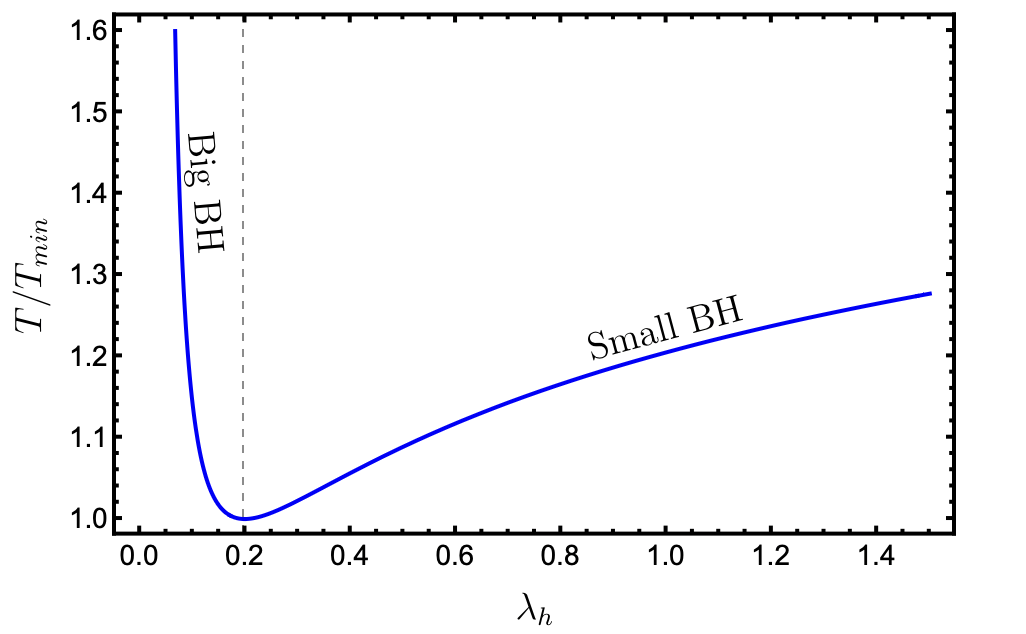}
    \caption{Temperature curve $T/T_{min}(\lambda_h)$.}
    \label{fig:temperature curve}
\end{figure}

We will work in the large $N_c$ limit, which means that the backreaction of the axion on the metric can be ignored. Then, at finite temperature, obtained after compactifying Euclidean time on a circle of radius $1/T$ (i.e. $\tau\sim\tau+1/T$), we have two solutions:
\begin{itemize}
    \item \textbf{Thermal gas (TG) solution:} The metric has the same form as in the zero-temperature solution, with the Euclidean time compactified
    \begin{equation}\label{thermal gas metric}
        ds^2=b_0(r)^2(dr^2+d\tau^2+dx_mdx^m), \quad \Phi=\Phi_0(r),
    \end{equation}
    where $r$ is a radial coordinate and $r\rightarrow0$ represents the AdS boundary. This solution describes thermal excitations around the vacuum represented by the zero-temperature solution. That being said, it represents the confined phase of the gauge theory.
    \item  \textbf{Black hole (BH) solutions}: These have the form
    \begin{equation}
        ds^2=b(r)^2\left[\frac{dr^2}{f(r)}+f(r)d\tau^2+dx_mdx^m\right], \quad \Phi=\Phi(r),
    \end{equation}
    and the presence of a solution with a regular horizon is characterized by
    \begin{equation}
        f(r_h)=0, \quad \dot{f}(r_h)<0,
    \end{equation}
    where the dot represents the derivative with respect to $r$. The horizon is a regular surface if the Euclidean time is identified as
    \begin{equation}
        \tau\sim \tau +\frac{4\pi}{\lvert \dot{f}(r_h)\rvert},
    \end{equation}
    and the temperature is given by the usual Hawking temperature
    \begin{equation}
     T=\frac{\lvert \dot{f}(r_h)\rvert}{4\pi}.
    \end{equation}
    This geometry corresponds to a deconfined phase in the dual gauge theory since the confining string tension is given by \cite{Kinar:1998vq}
    \begin{equation}
        \min_r(\sqrt{g_{\tau\tau}(r)g_{xx}(r)})\sim \sqrt{f(r_h)}=0.
    \end{equation}
\end{itemize}
\begin{figure}[t!]
    \centering
    \includegraphics[width=0.865\linewidth]{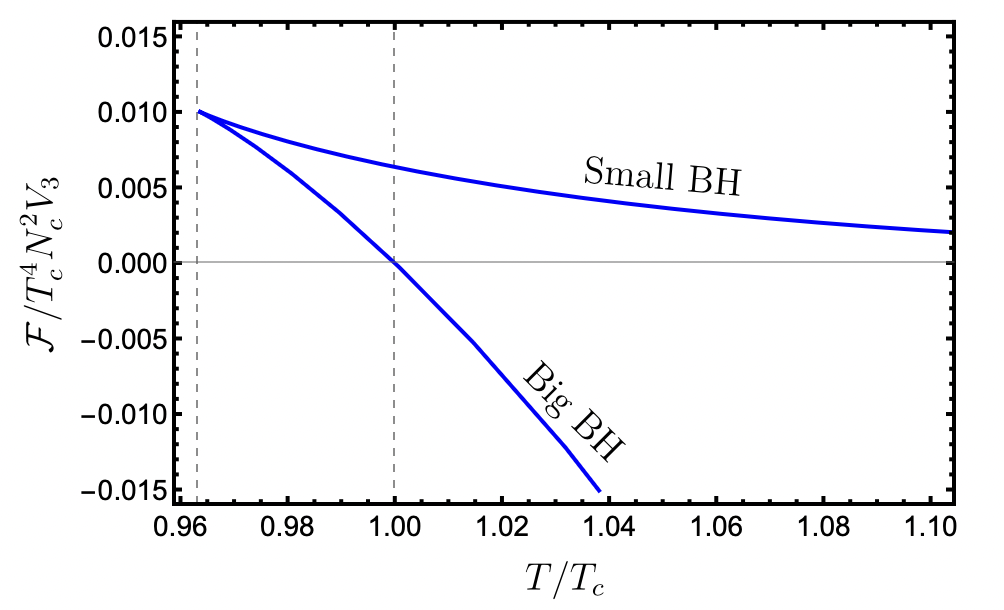}
    \caption{Free energy diagram of the $SU(N_c)$ YM theory without the $\theta$-term.}
    \label{fig:free energy curve}
\end{figure}
Since the solutions need to approach AdS asymptotically as $r\rightarrow 0$, in this limit we must have $b(r)\sim \ell/r$ and $f(r)\rightarrow 1$ up to logarithmic corrections. The typical solutions for $b(r)$ and $f(r)$, hence for the temperature $T$, are obtained numerically.\footnote{In this case, it appears easier to work with scalar variables $X=\dot{\Phi}/(3\dot{A})$ and $Y=\dot{g}/(4\dot{A})$, where $b(r)=e^{A(r)}$. This procedure is summarized in the Appendix of \cite{Gursoy:2008za}.} The holographic coordinate $r$ and the dilaton potential $V(\lambda)$ are in one-to-one correspondence with the running energy scale and the $\beta$-function of the dual gauge theory, respectively \cite{Gursoy:2007cb,Gursoy:2008za}. Thus, intuitively, and as shown more precisely in \cite{Gursoy:2007cb,Gursoy:2007er} this means that one can use the dilaton $\lambda$ as a coordinate to describe gravitational solutions. The temperature curve of the BH as a function of the dilaton value at the horizon $\lambda_h$ is shown in Figure \ref{fig:temperature curve}. One observes the existence of a minimal temperature $T_{min}$ ($\lambda_{h,min}\approx0.2$), below which the deconfined phase does not exist and the phase transition must be completed. For BHs with $\lambda_h$ smaller (larger) than $\lambda_{h,min}$, the horizon is closer to (further from) the AdS boundary and we refer to these as big (small) BHs.

The phase diagram is determined from the free energy curve. Namely, we look at the free energy difference between the black hole and the thermal gas solution 
\begin{equation}\label{free energy equation}
    \mathcal{F}=T(S_{BH}-S_{TG}),
\end{equation}
i.e. between the deconfined and confined phases. The action is generally divergent near the UV boundary $r=0$ and a cut-off is placed at $r=\epsilon$ or, equivalently, at some $\lambda=\lambda_0$. Taking the difference in \eqref{free energy equation}, we avoid the explicit computation of counterterms \cite{Gursoy:2008za}. The phase diagram can also be computed with the entropy
\begin{equation}
\mathcal{F}(\lambda_h)=\int_{\lambda_h}^\infty s(\tilde{\lambda}_h)\frac{dT}{d\tilde{\lambda}_h}d\tilde{\lambda}_h.
\end{equation}
The entropy is given by
\begin{equation}
    s(\lambda_h)=\frac{\text{Area}}{4G_5}=4\pi M_{Pl}^3N_c^2 V_3\ b^3(\lambda_h),
\end{equation}
where $V_3$ is the volume of the three-dimensional space spanned by the coordinates $x^m$. The typical phase diagram that one obtains is shown in Figure \ref{fig:free energy curve}. One can see the existence of a first-order phase transition at the critical temperature $T_c$. Moreover, the big BH represents the actual deconfined phase of the gauge theory, whereas the small BH shall be interpreted as the spinodal unstable phase, which in the case of large $N_{c}$ asymptotes to the confined phase at $\lambda_{h}\rightarrow \infty\,.$

Note that this diagram was obtained without including the axion contribution to the free energy. In what follows, we discuss the axion profile in these backgrounds and its contribution to the phase diagram.
\subsection{Axion backgound at zero and finite Temperature}
Even though we are working in the approximation where the axion does not backreact on the metric, it can still obtain a non-trivial background profile. Following \cite{Gursoy:2007er,Gursoy:2008za}, we will present the axion background at zero temperature and then extend it to non-zero temperature. As already mentioned, the only solution at zero temperature is the vacuum one, with the metric given in \eqref{thermal gas metric} and the Euclidean time uncompactified. The axion equation of motion is given by
\begin{equation}
    \ddot{a}+\left(3\frac{\dot{b}}{b}+\partial_{\lambda}\log Z(\lambda) \dot{\lambda}\right)\dot{a}=0.
\end{equation}
The equation above has two independent solutions
\begin{equation}
    a(r)=\theta_{UV}+2\pi k+ Ca_1(r), \quad k\in\mathbb{Z}, \,\,\,\theta_{UV}\in[0,2\pi\rangle
\end{equation}
where $2\pi k$ was added to account for the $2\pi$ periodicity of the $\theta$-angle in the UV. The non-trivial solution is given by
\begin{equation}
    a_1(r)=\int_0^r\frac{d\tilde{r}}{\ell}\frac{1}{b^3(\tilde{r})Z(\lambda)}
\end{equation}
and $C$ is an integration constant proportional to the expectation value of the instanton density 
\begin{equation}\label{Integration constant C}
    C=Z_0 \ell^4\frac{\langle F\wedge F\rangle}{32\pi^2}.
\end{equation}
One can interpret the non-trivial axion profile as the running $\theta$-angle, but this statement should be taken with some care. As the instanton density operator $\langle F\wedge F\rangle$ does not receive divergent UV corrections, the $\theta$-angle does not get renormalized \cite{Vicari:2008jw}. Instead, the point of view adopted in \cite{Gursoy:2007er} is that the axion profile captures the non-perturbative corrections to the $\theta$-angle in the IR.

One can calculate the $\theta$-dependent vacuum energy
\begin{equation}
    \mathcal{E}(\theta_{UV})=\frac{M_{Pl}^3}{2}\int dr\sqrt{g}Z(\Phi)(\partial a)^2.
\end{equation}
Using the above results, one can easily show that this is equal to \cite{Gursoy:2007er}
\begin{equation}\label{Theta dependent vacuum energy}
\mathcal{E}(\theta_{UV})=\frac{1}{2}\chi \ \min_k(\theta_{UV}+2\pi k)^2,
\end{equation}
where we had identified the topological susceptibility $\chi$ as
\begin{equation}
    \chi=\frac{M_{Pl}^3}{\int_0^{\infty}\frac{dr}{b^3(r)Z(\lambda)}}.
\end{equation}
and imposed the following IR boundary condition 
\begin{equation}
    \lim_{r\rightarrow\infty}a(r)=0.
\end{equation}
This boundary condition was adopted in \cite{Gursoy:2007er} motivated by the observation that, in the spirit of holography, physical quantities are determined from the data at a single boundary. Moreover, it is motivated in certain string-theory models where the axion is identified with a Wilson loop on a disc \cite{Witten:1998uka}. As the disc shrinks to zero size in the IR, the axion profile $a(r)$ should also vanish there.
\begin{figure}[!t]
    \centering
    \includegraphics[width=1\linewidth]{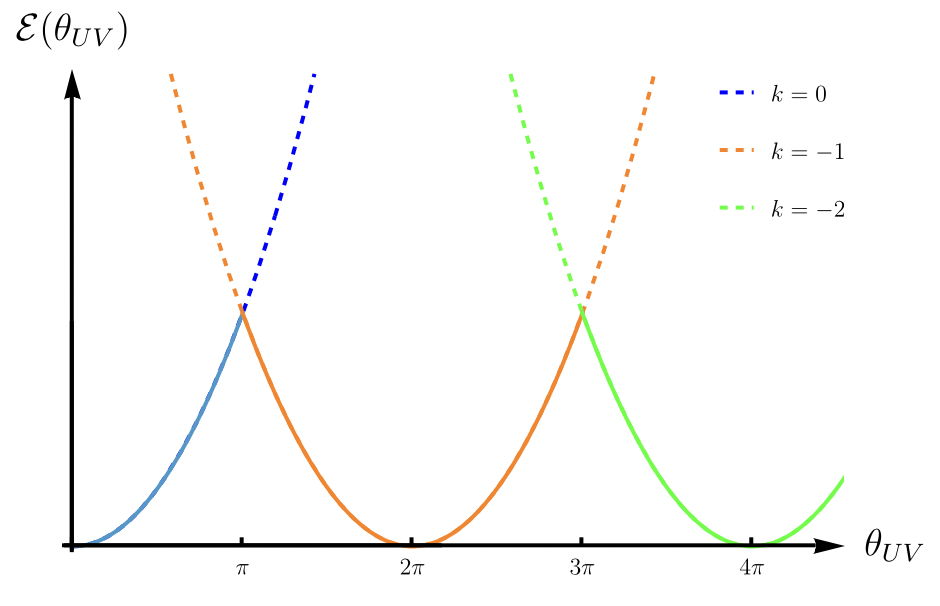}
    \caption{Illustration of the branch structure of the Yang-Mills vacuum. Solid lines of different colors represent the true vacuum for that value of $\theta_{UV}$, whereas dashed lines correspond to metastable vacua.}
    \label{fig:branches}
\end{figure}

From \eqref{Theta dependent vacuum energy} we can see that the vacuum structure of Yang-Mills is special. There is a single global minimum labelled by $k=k_{min}$ which minimizes \eqref{Theta dependent vacuum energy} for some value of $\theta_{UV}$ and all other integers $k\neq k_{min}$ represent metastable vacua. As the value of $\theta_{UV}$ increases, the two neighbouring minima, $k_{min}$ and $k_{min}-1$, become degenerate at $\theta_{UV}=(1-2k_{min})\pi$. Increasing $\theta_{UV}$ further causes the stable ($k_{min}$) and the metastable vacuum ($k_{min}$-1) to change roles (see Figure \ref{fig:branches}). Such a vacuum structure was initially proposed in \cite{Witten:1998uka} as it preserves the $2\pi$-periodicity of $\theta$ at large $N_c$.

At finite temperature, the axion equation of motion changes by an additional term
\begin{equation}
    \ddot{a}+\left(3\frac{\dot{b}}{b}+\frac{\dot{f}}{f}+\partial_{\lambda}\log Z(\lambda) \dot{\lambda}\right)\dot{a}=0
\end{equation}
and this can be easily integrated as before to obtain
\begin{equation}\label{axion at finite T}
    a(r)=\theta_{UV}+2\pi k+C \int_0^r\frac{d\tilde{r}}{\ell}\frac{1}{f(\tilde{r}) b^3(\tilde{r})Z(\lambda)},
\end{equation}
where $C$ is again given by \eqref{Integration constant C}. In the thermal gas solution, the axion profile coincides with that of the zero-temperature background. In contrast, in the black hole background, since near the horizon $f(r)\sim \dot{f}(r_h)(r_h-r)$, the integral in \eqref{axion at finite T} will be logarithmically divergent ($b$ and $Z$ are both finite near the horizon \cite{Gursoy:2008za}). Since the action must be finite, we must impose that $C=0$ and that
\begin{equation}
    \langle F\wedge F\rangle_{\text{deconf}.}=0,
\end{equation}
as expected from lattice simulations at large $N_c$ \cite{Bonati:2013tt,Lucini:2004yh,Vicari:2008jw}. 

However, as we move the position of the horizon towards infinity $r_h\rightarrow\infty$, the exterior geometry is approaching the thermal gas solution. This limit is not reproduced within the axion profile if $C=0$. This suggests that, at larger values of $r_h$, one should properly include the backreaction from the axion to account for the susceptibility in the SBH branch, which should approach the thermal gas one as $r_h\rightarrow\infty$. We will discuss the regime in which this approximation breaks down and the onset of significant axion backreaction.
\section{Confinement Phase Transition with finite $\theta$}\label{Section: Confining PT at finite theta}
We have shown the free energy diagram in Figure \ref{fig:free energy curve}, which signals a first-order transition between the deconfined and confined phase of Yang-Mills. It is illustrative to estimate the maximum amount of supercooling we can get from such phase transitions. Introducing the supercooling parameter \cite{Agrawal:2025wvf,Agrawal:2025xul}
\begin{equation}
    \epsilon=1-\frac{T_n}{T_c},
\end{equation}
where $T_n$ is the nucleation temperature, we see that the maximum supercooling one can obtain is 
\begin{equation}
    \epsilon_{max}=1-\frac{T_{min}}{T_c}\approx 0.04.
\end{equation}
For small supercooling, the parameter $\beta/H\sim1/\epsilon^3$ is larger, which suppresses the GW signal \cite{Agrawal:2025xul}. In this section, we investigate the impact of the axion on the supercooling of confining phase transitions in Yang-Mills theories.

We compute the contribution of the axion sector to the free energy density. In the black hole phase, this contribution vanishes since the axion profile is constant, and therefore we focus on the thermal gas solution. The axion part of the Euclidean action is given by
\begin{equation}
S_{a}^{TG}=\frac{M_{Pl}^3}{2}\frac{V_3}{T}\int_{0}^{\infty} dr \ b_0(r)^3\ Z(\lambda)\ (\partial a)^2.
\end{equation}
The integral above is precisely the $\theta$--dependent vacuum energy computed previously and it is easy to see that 
\begin{equation}
    S_a^{TG}=\frac{1}{2}\frac{V_3}{T}\ \chi \min_k(\theta_{UV}+2\pi k)^2.
\end{equation}
Since we will later consider cosmological estimates, it seems convenient to set the energy scale in terms of the critical temperature at $\theta_{UV}=0$, i.e. $T_c^{\theta_{UV}=0}$. Since $Z(\lambda)$ is matched using the lattice data, we use the scaling from \cite{DElia:2012pvq}
\begin{equation}
    \chi\approx0.0221 \sigma_{conf}^2, \quad T_c^{\theta_{UV}=0}\approx0.597 \sqrt{\sigma_{conf}},
\end{equation}
where $\sigma_{conf}$ is the confining string tension. Using this, the properly normalized contribution to the free energy from the axion is
\begin{equation}
\begin{split}
    \frac{\mathcal{F}_a}{N_c^2V_3(T^{\theta_{UV}=0}_c)^4}&=- \frac{1}{2}\frac{(\theta_{UV}+2\pi k)^2}{N_c^2}\frac{\chi}{(T^{\theta_{UV}=0}_c)^4}\\
    &\approx -\frac{(\theta_{UV}+2\pi k)^2}{N_c^2} (0.09),
\end{split}
\end{equation}
where the minus sign comes from \eqref{free energy equation}. 

\begin{figure}[t!]
    \centering
    \includegraphics[width=1.0\linewidth]{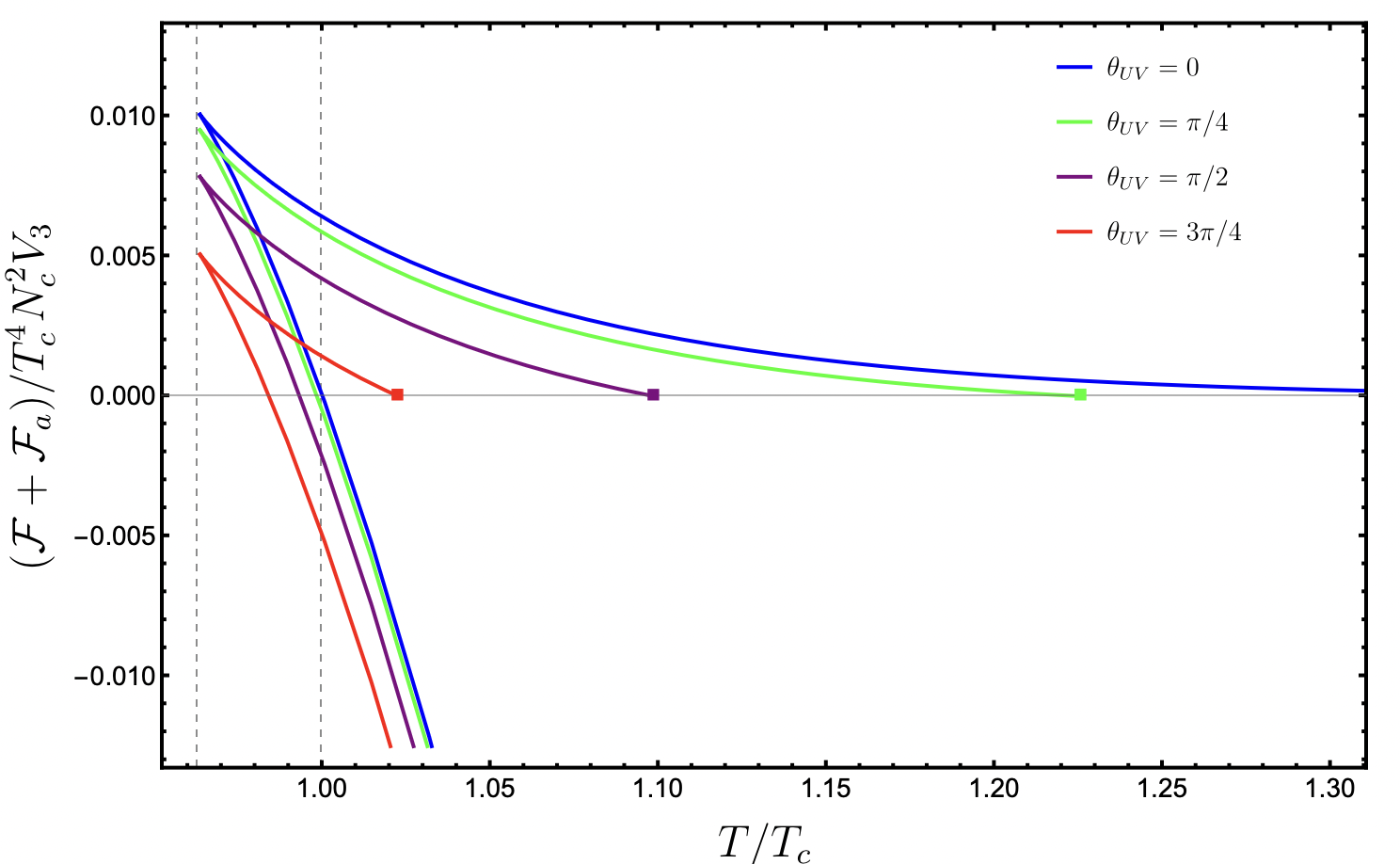}
    \caption{Free energy curve with the axion contribution for various values of $\theta_{UV}$. In this plot, we set $N_c=10$ and $k=0$. The squares indicate the value of $\lambda_h$ where the axion backreaction should be taken into account.}
    \label{fig:free energy with axions}
\end{figure}

What is the minimal $N_c$ for which the large $N_c$ approximation is valid and the axion backreaction can be ignored? The free energy curve without the axion contribution evaluated at $T_{min}$ is equal to
\begin{equation}
    \frac{\mathcal{F}}{(T^{\theta_{UV}=0}_c)^4N_c^2V_3}\left(\frac{T_{min}}{T^{\theta_{UV}=0}_c}\right)\approx0.01.
\end{equation}
One may ask what the minimal value of $N_c$ is such that the contribution from the axion is equal to 0.01, since in this case it would spoil the metastability of the deconfined phase. In this case, one would need to include the full backreaction from the axion. That being said, the minimal value of $N_c$ is given by
\begin{equation}
    (N_c)_{min}\approx (\theta_{UV}+2\pi k)\times 3
\end{equation}
Taking the $k=0$ branch to be stable, and as a reference value, $\theta_{UV}=\pi$, where it becomes metastable, we obtain $(N_c)_{min}\approx 9.4$. For some values of $\theta_{UV}$ and $N_c=10$ in the $k=0$ branch we find curves in Figure \ref{fig:free energy with axions}. 
Additionally, as discussed in the previous section, we identify the breakdown of our calculation along the SBH branch with the points where the free energy crosses zero (indicated by squares in Figure \ref{fig:free energy with axions}). This can be understood by interpreting the SBH as the sphaleron mediating tunneling to the confined phase. In this picture, we expect that including the full axion backreaction would smooth out the free energy curve, causing it to asymptotically approach zero along the SBH branch.

Since $\lambda_h$ can be interpreted as the order parameter of the phase transition \cite{Morgante:2022zvc}, these restrictions can also be understood from the perspective of the effective potential, i.e. $V_{\textrm{eff}}(\lambda_h,T)$. In the regimes discussed above, the contributions from the axion become comparable to the ones coming from the metric and the dilaton, requiring the full solution with the axion backreaction included. Furthermore, since the SBH corresponds to the local maximum in the effective potential, we expect the axion backreaction to become increasingly important as we vary $\lambda_h$ from the local maximum to infinity.

\begin{figure}[t!]
    \centering
    \includegraphics[width=0.9\linewidth]{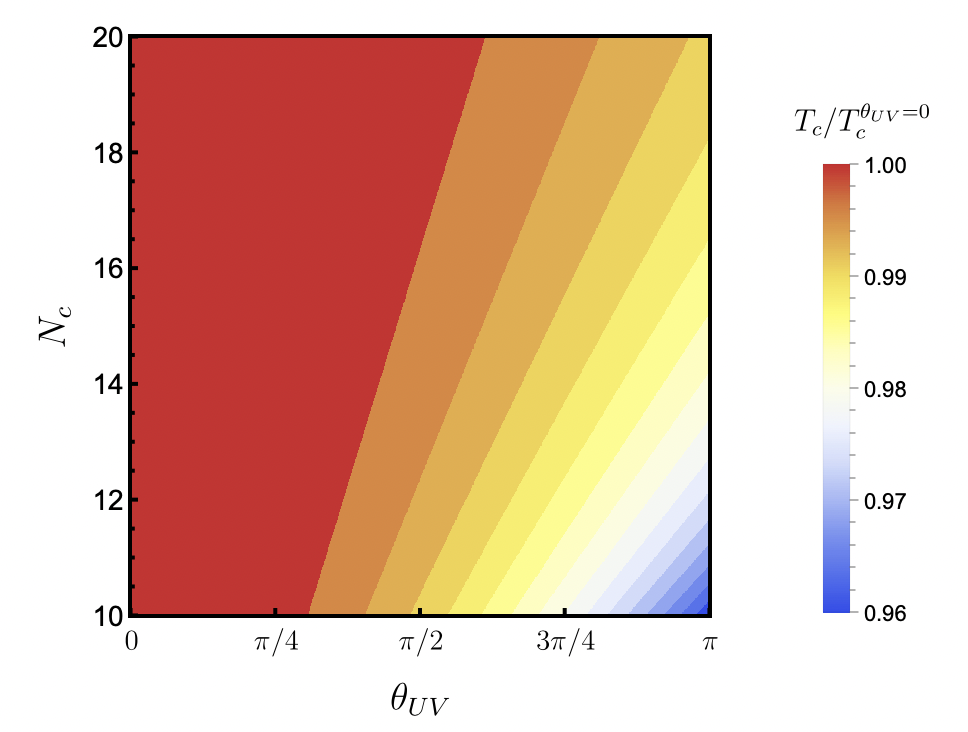}
    \caption{Contours of $T_c/T_c^{\theta_{UV}=0}$ in the $(N_c,\theta_{UV})$ plane. }
    \label{fig:Tc contours}
\end{figure}

From Figure \ref{fig:free energy with axions} we see that the addition of an axion reduces the amount of supercooling in the deconf./conf. phase transition of Yang-Mills theories, since the difference between $T_c$ and $T_{min}$ gets smaller and the GW spectrum gets more suppressed. The reduction in the critical temperature is, of course, in agreement with the lattice calculations \cite{DElia:2012pvq,Bonanno:2023hhp}. However, the notion of minimal temperature $T_{min}$, present in holography but harder to extract from lattice simulations, is crucial to estimate the maximum amount of supercooling. We show the contours of $T_c$ at finite $\theta\,$ and $N_{c}$ in Figure \ref{fig:Tc contours}.

In summary, it is clear from our discussion that a non-zero $\theta$-angle makes the PT less supercooled, making the prospects for GW signals worse.
In the next section, we will investigate the formation of domain walls due to the multi-branched structure of the YM vacuum.

\section{Domain wall production between different branches}\label{Section: DW production between different branches}

In the previous section, we discussed how the presence of the $\theta$-angle reduces the amount of supercooling in the confinement PT and, thus, the possible GW signal.
However, there is an additional effect due to the multi-branched structure of the YM vacuum. After the deconfinement/confinement phase transition in the early universe has completed, different $k$-vacua can be populated, producing domain walls across different patches in the universe. Since the vacua are not degenerate, domain wall annihilation will occur\footnote{Note that one can also tunnel between different branches via bubble nucleation \cite{Shifman:1998if,Sugeno:2025kwx}. The tunneling rate scales as $\Gamma \sim e^{-S}$, where $S\sim N_c^4$ and this process is suppressed at large $N_c$.}, and this is expected to leave a GW signal \cite{Babichev:2025stm,Notari:2025kqq}. 

\begin{figure}[t!]
    \centering
    \includegraphics[width=0.8\linewidth]{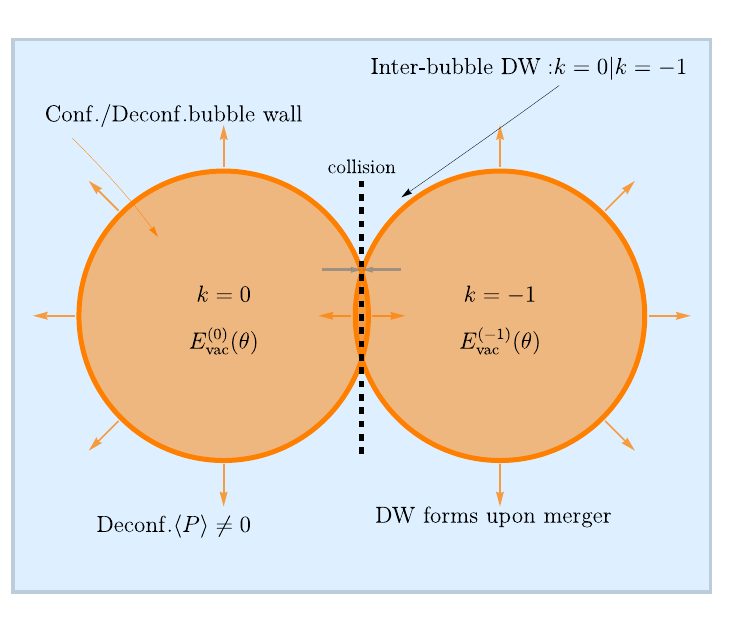}
    \caption{Illustration of bubble coalescence for the two bubbles of different $k$ branches, indicating that the domain walls can form within the collision region. $\langle P\rangle$ represents the VEV of the Polaykov loop.}
    \label{fig:two bubbles}
\end{figure}

\begin{figure}[t!]
    \centering
    \includegraphics[width=0.8\linewidth]{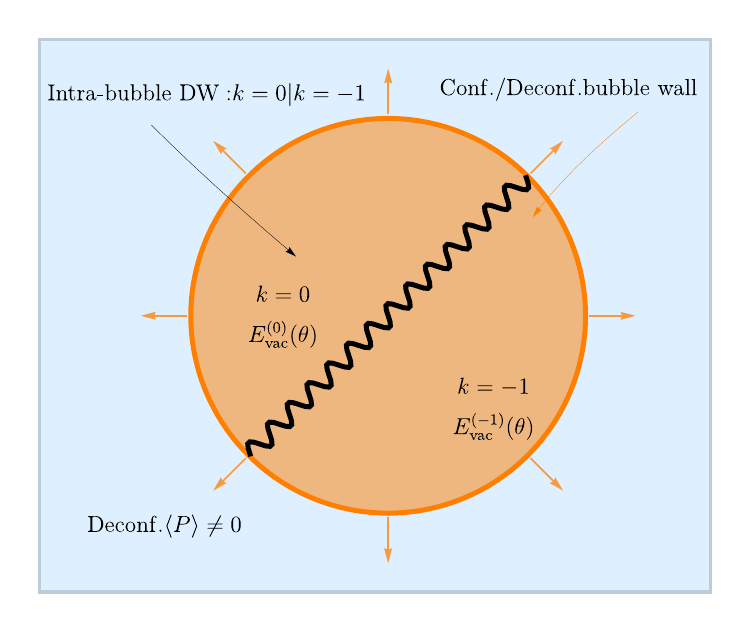}
    \caption{Illustration of a domain wall forming within a confining bubble. As the domain wall breaks the spherical symmetry of the bubble, these configurations are expected to be suppressed.}
    \label{fig:single}
\end{figure}

In this section, we discuss the conditions under which DWs can form and later annihilate.
We will consider for simplicity $\theta_{UV}\in[0,\pi\rangle$, so that the $k=0$ branch is stable and $k\neq0$ ones are metastable. During the phase transition, the interior of confining bubbles will contain a stable/metastable vacuum in some branch $k$ and domain walls will form at locations where the bubbles collide (Figure \ref{fig:two bubbles}). It can also happen that within the confining bubble, we have two vacua in different branches separated by a domain wall (Figure \ref{fig:single}). However, we expect nucleation of such configurations to be suppressed as they break the spherical symmetry within the bubble.

\subsection{Percolation theory}

Once a bubble is nucleated, the probability that its interior is in the $k$-th branch can be estimated as
\begin{equation}\label{bubble k probability general}
    p_k(\theta_{UV})=\frac{e^{-\Delta E_k/T}}{\sum_{j\in \mathbb{Z}}e^{-\Delta E_j/T}}.
\end{equation}
where $\Delta E_k= E_k- E_0$. This means that
\begin{equation}\label{eq:pk ratio}
    \frac{p_k(\theta_{UV})}{p_0(\theta_{UV})}=e^{-\Delta E_k/T}=\exp \left(-\frac{\chi}{T}(2k\pi)(\theta_{UV}+k\pi) V_0\right).
\end{equation}
where $V_0$ is given by the volume of the bubble at the moment of nucleation. Since the only scale at our disposal is the critical temperature $T_c$ (which is related to the strong coupling scale by $\Lambda\approx1.17 \,T_c$ \cite{Morgante:2022zvc}), we expect that the initial radius of the bubble will be $R_0\sim T_c^{-1}$%
and that the bubble volume is independent of $k$.%
\footnote{The $k$ dependence can be estimated in the thin-wall approximation \cite{Coleman:1977py, Callan:1977pt} by noticing that the bubble radius scales as the inverse of the energy difference of the two vacua. Thus we expect a $N_c^2$ suppressed correction of the form
\begin{equation}
    R_0\sim \frac{1}{\Lambda}\left(1+c\frac{(\theta_{UV}+2\pi k)^2}{N_c^2}\right)^{-1},
\end{equation}
where $c\sim\mathcal{O}(1)$.
}
The actual value can be obtained by solving the bounce equation. We find a typical value~\cite{Morgante:2026}
\begin{equation}
    R_0 \approx \frac{\xi}{T_c} \quad \text{with}\quad \xi = \mathcal{O}(10).
\end{equation}
and thus
\begin{equation}\label{ratio of probabilites k}
    \frac{p_k(\theta_{UV})}{p_0(\theta_{UV})}\approx\exp\left[-4.58\, \xi^3 \, k(k\pi+\theta_{UV})\right].
\end{equation}
Depending on the value of $\theta_{UV}$, there can be a finite probability of nucleating bubbles of the lowest-energy metastable branch $k=-1$. The nucleation of other metastable branches is significantly suppressed, and we will only consider domain walls between $k=0$ and $k=-1$ vacua.  Note that the critical temperature has disappeared from the expression in \eqref{ratio of probabilites k}, since all dimensionful quantities in the theory scale with the critical temperature.

\begin{figure}[t!]
    \centering
    \subfloat[$\epsilon_{\theta}=4\times10^{-4}$]{
        \includegraphics[width=0.22\textwidth]{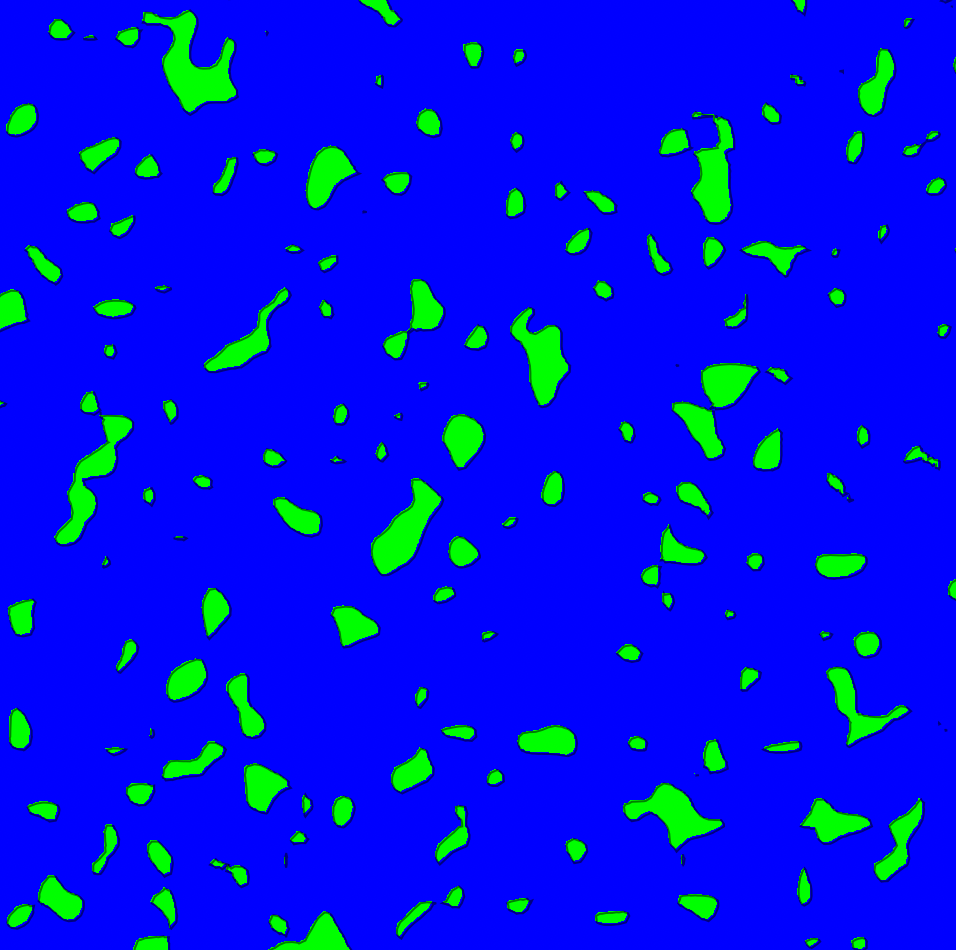}
        \label{fig:4104}
    }
    \hfill
    \subfloat[$\epsilon_{\theta}=5\times10^{-5}$]{
        \includegraphics[width=0.22\textwidth]{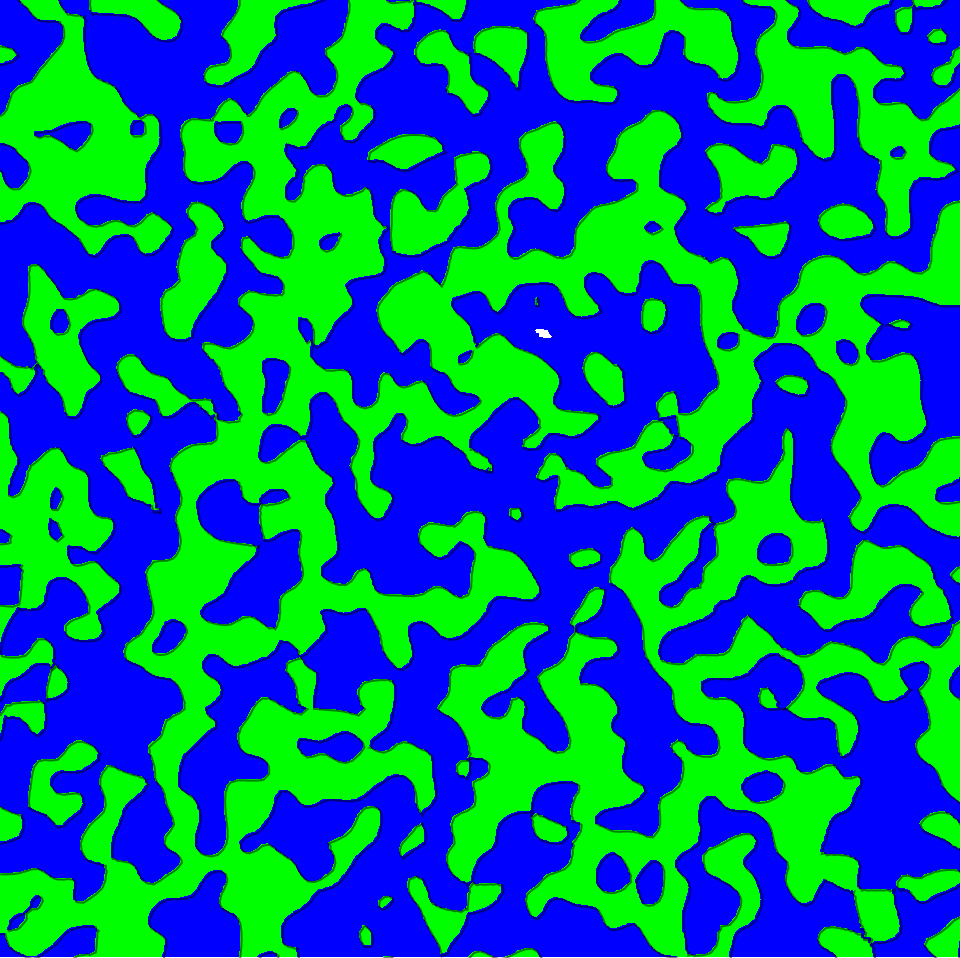}
        \label{fig:5105}
    }
    \caption{
    Formation of a domain-wall network initially depends on whether $\theta_{UV}$ satisifes \eqref{percolation bound on theta}: (a) $\theta_{UV}$ does not satisfy \eqref{percolation bound on theta} and the DW network does not form. (b) $\theta_{UV}$ satisfies \eqref{percolation bound on theta} and the network does form. The domains of $k=0$ and $k=-1$ vacua are shown in blue and green, respectively, and $\xi=10$ in this Figure.}
    \label{fig:DW network illustration}
\end{figure}

Although the probability for the nucleation of $k=-1$ bubbles can be finite, the question remains whether there is a significant number of $k=-1$ bubbles nucleating to form a domain wall network. We follow the reasoning given in \cite{Vilenkin:2000jqa}, which rephrases this problem in the language of percolation theory \cite{STAUFFER19791,Lalak:1993bp,Saberi_2015}. In other words, we need to ask whether one can achieve connectivity of a single vacuum domain across large scales. 
We first assume that bubbles grow until they fill the universe while satisfying Eq.~(\ref{eq:pk ratio}).
The critical probability for percolation in 3D is given by $p_c=0.31$ \cite{Vilenkin:2000jqa}. Hence, if $p_k(\theta_{UV})>p_c$ we can have a long-connected vacuum domain in branch $k$. In the opposite case, if $p_k(\theta_{UV})<p_c$, we can have only finite-volume domains which collapse under their own surface tension. Applying the percolation bound for the $k=-1$ branch, we can place a lower limit on $\theta_{UV}$ for which we achieve a DW network
\begin{equation}\label{percolation bound on theta}
    \pi-\theta_{UV}\equiv\epsilon_\theta\lesssim \frac{5.6\times10^{-2}}{\xi^3}\pi,
\end{equation}
where we introduced the parameter $\epsilon_{\theta}$ that indicates how close $\theta_{UV}$ is to $\pi$.
Thus, for $\theta_{UV}\to\pi$, after the confining bubbles percolate and the phase transition completes, a domain wall network can form between them, as illustrated in Figure \ref{fig:DW network illustration}. The typical domain length scale is given by the bubble radius at the end of the PT
\begin{equation}\label{bubble separation}
    R_*\approx (8\pi)^{1/3}v_w\left(\frac{1}{\beta/H}\right) \ \frac{1}{H}\ll\frac{1}{H},
\end{equation}
for low supercooling (typical values of $\beta/H$ for the confinement PT are $\mathcal{O}(10^5)$ \cite{Morgante:2022zvc}), where $v_w$ is the wall velocity.

Note that our criteria from percolation theory should be considered as an estimate, since the distribution of bubbles at the time of percolation and vacuum branch assignment are generated dynamically. Moreover, differences in bubble expansion velocities have been ignored. However, due to the large number of bubbles within a Hubble volume, i.e. large $\beta/H$, the percolation theory estimate \eqref{percolation bound on theta} still provides us with a reasonable indication for which values of $\theta$ the DW network can form.

\medskip

For now we have mainly considered the possibility of forming the DW network within the picture provided by percolation theory. We will now elaborate on the dynamics of bubble coalescence and DW formation.

\subsection{Dynamics of bubble coalescence}

\subsubsection{Before Bubble Coalescence}
 The duration of the phase transition, from the moment of nucleation ($T_{n}$) to that of bubble percolation ($T_{p}$) is given by $\beta^{-1}$, where \cite{Enqvist:1991xw}
 \begin{equation}
     \beta=HT\frac{dS_b}{dT} \bigg \rvert_{T=T_p}
 \end{equation} 
and $S_{b}$ is the bounce action. The ratio $\beta/H$ determines the average bubble separation $R_*$ in \eqref{bubble separation}. 
Initially, the bubble is nucleated in a region of space with either $k=0$ or $k=-1$ and begins to expand to eventually coalesce with another bubble nucleated nearby, possibly with a different $k$. This is happening throughout the Hubble volume due to the large $\beta/H$. 
As a confining bubble nucleates and begins to expand, the topological susceptibility $\chi$ inside this region rapidly approaches its confined asymptotical value on a time scale much shorter than $\beta^{-1}$ at percolation. This can be understood from the fact that localized equilibration in strongly coupled plasma dynamics is of the order $\tau_{\rm eq}\sim \mathcal{O}(1/T_c)$.

The confinement FOPT is most likely a strong deflagration, characterized by relatively large transition strengths, $\alpha \gtrsim 0.1$, and subsonic bubble wall velocities, $v_w \leq c_s(T_n)$. In this case, the substantial latent heat released can locally reheat the region surrounding the expanding true-vacuum bubbles above $T_c$, leading to the formation of remnant false-vacuum droplets and temporarily interrupting bubble coalescence \cite{Cutting:2022zgd, Gouttenoire:2023roe, Sanchez-Garitaonandia:2023zqz}.
One can estimate the locally reheated temperature as $T_{\rm re,loc}\sim (1+\alpha)^{1/4}T_{n}\,$, with $\alpha$ the amount of latent heat released. We have that $\alpha \sim 0.3-0.4$~\cite{Morgante:2022zvc, Morgante:2026} and $T_{n}\sim 0.99T_{c}$, hence $T_{\rm re,loc} \simeq 1.06\,T_c\,.$
This estimate is only valid right at the moment of nucleation and can be compared with the simulations conducted in Ref.~\cite{Cutting:2019zws} (see their Fig.~11), where they find that at bubble nucleation, the local reheated region has heated up to around 15\% above the nucleation temperature.%
\footnote{The effect is more prominent when the bubble wall travels slowly. In such transitions, the GW spectrum can be further suppressed by the fact that the kinetic energy is converted into heat in the pocket regions~\cite{Cutting:2019zws}.}
The key point is that false-vacuum droplets can still be present even after bubble coalescence, which can delay the formation of the DW network.

Therefore, in the subsequent subsection we want to elaborate on the dynamics as the initial bubble coalescence proceeds. We will discuss the subsequent evolution relying on macroscopic hydrodynamic relaxation toward equilibrium and use linearized hydrodynamics for qualitative estimates.   


\subsubsection{After Bubble Coalescence}

In this section, we present the timescales relevant for the formation of the DW network after bubble coalescence. We use insights from the following collection of references regarding holographic thermalization, planar AdS/Asymptotic AdS wall collisions, and strongly coupled equilibration dynamics \cite{Balasubramanian:2011ur, Chesler:2011ds,Heller:2011ju, Heller:2012je,Berges:2020fwq, Buchel:2015saa,Buchel:2011wx, Gursoy:2009kk,Janik:2015waa, Janik:2014kfa, Attems:2016tby, Attems:2017zam}. We will summarize our findings below and highlight some more details in Appendix \ref{Relevant timescales}. 

First, we will look at the set of timescales of the local collision region itself, which has a size $\ell_{\rm local}\sim \mathcal{O}(1/T_c)\sim \mathcal{O}(1/\Lambda)$. This will give us the timescale for the emergence of local DW segments within the collision region. Following that, we will characterize the timescale when the thermal fluctuations of the system on larger scales $R_{*}\gg \ell_{\rm local}$ become damped, denoted by $\tau_{\rm damp}$, which estimates the initial emergence of large connected DW network regions.

During bubble coalescence, the collision regions are initially driven far from equilibrium and undergo substantial local reheating, whereas remnant false-vacuum droplets still persist alongside the forming network. The post-collision system can be characterized by the effective temperature $T_{\rm eff}(\tau,x)$, which immediately after the collision satisfies $T_{\rm eff}(\tau,x)>T_p$. We remind the reader that for strongly coupled FOPTs with little supercooling $T_p\sim T_c$. In order for domain walls to start to form locally, the remnant droplets must also cool below $T_c$, so that the surface tension of the domain wall is well-defined. Then the relevant local timescales are the following:
\begin{itemize}
\item At $\tau=\tau_{\rm hydro}$ the system can be described by viscous hydrodynamics, even though the system is still out of equilibrium;
\item At $\tau_{\rm eq}$ the system approaches an effective local thermal equilibrium on a length scale $\ell_\mathrm{local}$;
\item $\tau^{\rm loc}_{\rm cool}$ is the time required for the locally reheated region to satisfy $T_{\rm eff}(\tau,x)<T_c$.
\item Finally, at $\tau_{\rm micro}$ non-hydrodynamical modes (large momenta) on microscopic scales thermalize, as reflected in the evolution of two-point correlation functions such as $\langle \mathcal{O}_{\phi}\mathcal{O}_{\phi}\rangle\,$ or $ \langle T_{\mu\nu}T^{\mu\nu}\rangle\,,$ towards their thermal equilibrium values.
\end{itemize}
Once $\tau\sim \tau_{\rm hydro}$, the effective temperature of the system $T_{\rm eff}$ has already started settling towards $T_{c}$ so that the subsequent evolution thereafter takes place with an effective temperature $T_{\rm eff}\sim T_{c}\,$~\cite{Heller:2011ju,Heller:2012je}.  
The relevant local relaxation scales obey the hierarchy\footnote{This hierarchy of timescales persists both in conformal and non-conformal strongly-coupled plasmas, whereas the actual values differ by $\mathcal{O}(1)$  numbers, i.e. $\tau^{\rm nonCFT}/\tau^{\rm CFT}\sim \mathcal{O}(1)$. For details, see Appendix \ref{Relevant timescales} or \cite{Berges:2020fwq,Attems:2016tby,Attems:2017zam,Attems:2018gou,Janik:2015waa}\,.}
\begin{equation}
\tau_{\rm hydro}\lesssim 
\tau_{\rm eq}\sim \tau_{\rm micro} .   
\end{equation}
All of these scales are of the order ${\cal O}(1/T_c)\sim \mathcal{O}(1/\Lambda)$ in strongly coupled plasmas \cite{Heller:2011ju,Heller:2012je}. 

The initial formation of local DW segments can occur when, at least within the effective LTE regime, the associated temperature in a region drops below $T_{c}$. The local DW segment formation time can therefore be estimated as
\begin{equation}
\tau_{\rm DW}^{\rm loc}
\simeq
\max(\tau_{\rm eq},\tau^{\rm loc}_{\rm cool}).
\end{equation}
Thus, $\tau_{\rm DW}^{\rm loc}$ refers to the DW formation within the local post-collision
region, whose length scale is set by the wall thickness $\ell_{\rm DW}\sim \ell_{\rm local}$.

\bigskip

The next question is when these local DW segments can merge into a connected network. Even after approaching LTE within the collision region of two bubbles, large-scale temperature inhomogeneities persist over the mean bubble separation $R_*$.
These fluctuations equilibrate on a timescale $\tau_\mathrm{damp}$, which should be shorter than a Hubble time to allow the formation of the DW network, i.e. $\tau_{\rm damp}H< 1$. We estimate this damping timescale using linearized hydrodynamics.
The amplitude of long-wavelength sound waves goes as
\begin{equation}
    \mathcal{A}\sim e^{-i  \omega(k) t}\,,
\end{equation}
where the dispersion relation for $\omega(k)$ is given by 
\begin{equation}\label{omega expansion}
    \omega(k)= c_{s} k - \frac{i}{2}\Gamma_s k^{2} + \mathcal{O}(k^{3})\,,
\end{equation}
and we have chosen the branch with positive real part \cite{Bea:2021zol,Jeon:2015dfa,Attems:2019yqn}. The negative imaginary part describes viscous attenuation. The expansion of Eq. \eqref{omega expansion} is only valid in the hydrodynamical regime at low momenta. The sound attenuation constant $\Gamma_s$  is given by  
\begin{equation}
    \Gamma_s = \frac{1}{T}\left(\frac{4\eta(T)}{3 s} + \frac{\zeta(T)}{s}\right)\,,
\end{equation}
where $\eta$ is the shear viscosity and $\zeta$ is the bulk viscosity of the plasma.
Hence, the attenuation of sound waves/thermal fluctuations evolves as
\begin{equation}
    A(t)\simeq A_{0}e^{-\frac{1}{2}k^{2}\Gamma_{s}t}\,.
\end{equation}
For the damping time $\tau_{\rm damp}\sim 2/(k^{2}\Gamma_{s})$ over length scales of order $k^{-1}\sim R_{*}$ we get
\begin{equation}
    \tau_{\rm damp} \simeq \mathcal{O}(1) \frac{R_{*}^{2}}{\Gamma_{s}}\,,
\end{equation}
where $T\simeq T_{c}$. From Appendix \ref{Relevant timescales} we know that in IHQCD
\begin{equation}
\Gamma_{s} =  \frac{1}{sT_{c}}\left(\frac{7}{3}\eta\right)\,,
\end{equation}
where $\zeta/s \sim \eta/s = 1/(4\pi)$, at $T\simeq T_{c}$.
Additionally, using $R_{*}H\simeq 10^{-5}v_{w}$, we find that the thermal damping timescale is roughly given by 
\begin{equation}
    H \tau_{\rm damp} \sim \frac{12\pi 10^{-10}v_{w}^{2}}{7} \left(\frac{T_{c}}{H} \right)\,, 
\end{equation}
where, assuming radiation domination, we have
\begin{equation}
    H(T)\simeq \frac{\pi}{3}\left(\frac{g_*(T)}{10}\right)^{1/2}\frac{T^2}{M_\mathrm{Pl,4d}}
\end{equation}
and $M_\mathrm{Pl,4d}=2.4\times10^{18}\text{ GeV}$ is the reduced Planck mass, which is not to be confused with $M_\mathrm{Pl}$ used in holographic calculations above. 
Ultimately, we find that the time duration for damping of local thermal fluctuations is given by%
\footnote{At large $N_c$ and in the deconfined phase, our dark YM sector significantly contributes to $g_*(T)$. Here, the relevant scenario occurs after the confining PT has been completed, in which the spectrum consists of massive glueballs which do not contribute to $g_*(T)$.}
(using $g_*(T_c)\approx100$)  
\begin{equation}
    H\tau_{\rm damp} \sim \frac{12\pi}{7}v_{w}^{2}10^{-10}\left( \frac{M_{Pl,4d}}{T_{c}} \right)\,.
\end{equation} 
The condition for fast damping of thermal fluctuations $\tau_{\rm damp} H\leq1$ gives us a lower bound on $T_{c}$ 
\begin{equation}\label{bound on Tc from thermal fluctuations}
    T_{c}\gtrsim 1.3v_{w}^{2} \times 10^9 \text{ GeV} \,.
\end{equation} 
As long as this bound on $T_{c}$ is satisfied, the thermal fluctuations of size $R_{*}$ produced in the surrounding plasma during bubble coalescence are damped in less than a Hubble time and will not obstruct the formation of long connected regions.  Hence, the DW network is formed when 
\begin{equation}
    \tau_{\rm DW}^{\rm net}\sim \tau_{\rm damp}.
\end{equation} 
The scenarios presented in Ref.~\cite{Morgante:2022zvc}, with $T_c = 50\,\mathrm{MeV}, \,100\,\mathrm{GeV}$, and $v_w\sim 0.01$, violate Eq.~(\ref{bound on Tc from thermal fluctuations}), leading to a delay in DW formation, as we will now discuss.

If Eq.~(\ref{bound on Tc from thermal fluctuations}) is violated and $\tau_{\rm damp}H>1$, the connection of local DW segments is delayed due to large temperature fluctuations present in the plasma.  There are two important consequences. On the one hand,  as the universe expands and cools, the typical length of DWs, $R_{*}$, has increased, and the initial population of DWs would be less overdense.  On the other hand, the network may never form, since the potential bias $\delta V $ breaking the degeneracy of the vacua may already dominate the surface tension $\sigma$ leading to annihilation. In the following, we will assume that we are in the scenario where \eqref{bound on Tc from thermal fluctuations} is satisfied and the domain wall formation proceeds quickly.

\subsection{Scaling regime}
As we assume that the DW network forms quickly after the completion of the phase transition, the initial correlation length of the network is roughly given by $R_*$. This length scale will evolve until the scaling regime is reached. The scaling regime is an equilibrium attractor-like solution in which the correlation length of the network becomes comparable to the Hubble scale \cite{Press:1989yh, Kawano:1989mw, Avelino:2005kn, Leite:2011sc, Dankovsky:2024zvs, Blasi:2025tmn}.
During their evolution, the energy density of domain walls is given by
\begin{equation}
    \rho_{DW}\approx\frac{\sigma L^2}{L^3}\approx\sigma L^{–1},
\end{equation}
where $L$ is the length scale at that instance in the evolution and $\sigma$ is the surface tension of the walls. However, as discussed previously, the vacua considered here are not degenerate and there is a potential bias $\delta V$ between them. This leads to annihilation, and DWs will annihilate when 
\begin{equation}\label{annihilation condition}
    \rho_{DW}=\delta V.
\end{equation}
Since the GW spectrum is the strongest if the network has been in the scaling regime for some time \cite{Babichev:2025stm, Cyr:2025nzf}, we need to check whether the scaling regime can be reached at all in our framework.

To estimate whether our network can evolve to the scaling regime, one needs to find the surface tension of the domain walls $\sigma$. As this is not straightforward, we draw some intuition from the considerations in $\mathcal{N}=1$ SYM \cite{Gabadadze:2000vw, Dvali:1996xe, Shifman:1998if, Dvali:1998ms}. In that theory, gaugino condensation spontaneously breaks the global $\mathbb{Z}_{2N_c}$ (the non-anomalous subgroup of $U(1)_R$) to $\mathbb{Z}_2$, leaving behind $N_c$ degenerate vacua labeled by an integer $k$. The tension of the domain wall interpolating between the $k$-th and $k^\prime$-th vacuum was found to be \cite{Dvali:1998ms, Shifman:1998if, Dvali:1996xe}
\begin{equation}
    \sigma_{k,k^\prime}=\frac{N_c^2}{4\pi^2}\left\lvert\sin\left(\frac{\pi(k-k^\prime)}{N_c}\right)\right\rvert\Lambda^3.
\end{equation}
SUSY can be softly broken by adding a small gaugino mass $m_g$, which breaks the degeneracy between the vacua \cite{Shifman:1998if}. The SUSY-breaking term does not change the surface tension as long as $m_g\ll\Lambda$. For $m_g\gtrsim \Lambda$, the tension will change by an $\mathcal{O}(1)$ factor, but the overall scaling with $N_c$ and $\Lambda$ will not change. As there is no quantitative prediction of this surface tension computed with IHQCD, we adopt the SUSY result in our GW estimates.

As before, we will only consider domain walls separating the two lowest branches, i.e. $k=0$ and $k=-1$. The domain-wall tension is given by
\begin{equation}
    \sigma_{0,-1}\simeq\frac{N_c}{4\pi}\Lambda^3,
\end{equation}
in the large $N_c$ approximation. From Eq.~\eqref{Theta dependent vacuum energy}, we can easily calculate the bias term for $\theta_\mathrm{UV}\to\pi^-$
\begin{equation}
    \delta V=2\pi(\pi-\theta_{UV})\chi=2\pi \epsilon_\theta\chi.
\end{equation}

So, let us estimate whether our network evolves into the scaling regime or gets annihilated immediately after the completion of the phase transition ($T\approx T_c$). Since the length scale of domains after the phase transition is roughly given by $R_*$, the domain walls will survive annihilation if the surface tension wins over the bias
\begin{equation}
    \sigma R_*^{-1}\gtrsim\delta V,
\end{equation}
where $R_*\sim 10^{-5} H(T_c)^{-1}$ for typical parameters considered in \cite{Morgante:2022zvc}. 
Using the lattice inputs as before, we find
\begin{equation}\label{bound on Tc}
    T_c\gtrsim \frac{\epsilon_{\theta}}{N_c}\ 2\cdot 10^{14}\text{ GeV}.
\end{equation}
If the critical temperature violates the inequality in \eqref{bound on Tc}, the domain wall annihilation will proceed immediately. This inequality is generally violated for the critical temperatures considered in \cite{Morgante:2022zvc}, unless $N_c$ is insanely large or $\epsilon_{\theta}\sim10^{-12}$. For larger $N_{c}$, the range of validity in our calculation improves and may be the most suitable parameter to vary over a wider range. 

We assume that the inequality in \eqref{bound on Tc} is satisfied and that the domain walls can reach the scaling regime. This raises the question of how long it takes for the formed 
domain wall network to reach the scaling regime. In simulations of domain wall networks, carried 
out mainly in weakly coupled theories, it takes the network roughly $3$--$7$ Hubble 
times to reach the attractor scaling solution~\cite{Notari:2025kqq}. 
As discussed previously, the upshot here is that local microscopic equilibration in strongly coupled theories is a rapid process with relaxation timescales of order $\tau \sim T_c^{-1}$, thus removing one source of dynamical uncertainty. Hence, one could anticipate that approaching scaling in these theories would be at least equally fast as in weakly-coupled theories, if not slightly faster.  
However, since the approach to scaling regime is a dynamical phenomenon, simulations are necessary to accurately quantify the claims above.

\subsection{Summary}
As the discussion on DWs was rather lengthy, we would like to provide a small summary with the most relevant results. During the phase transition of the $SU(N_c)$ YM theory, confining bubbles can occupy different vacuum branches which can lead to a DW network under certain conditions once the phase transition has completed. These conditions can be listed below by their chronological relevance:
\begin{itemize}
    \item The bound on $\theta_{UV}$ (or $\epsilon_{\theta}$) from percolation theory (Eq. \eqref{percolation bound on theta}) tells us whether there are enough bubbles with vacuum branch index $k$ nucleating to form a DW network at all.
    \item The bound on $T_c$ in Eq.~\eqref{bound on Tc from thermal fluctuations} comes from the requirement that the thermal fluctuations relax within a Hubble time. If this is not satisfied, the formation of a global DW network will be delayed, suppressing the possible GW signal, if a scaling network forms.
    \item The condition on $T_c$ in Eq. \eqref{bound on Tc} determines whether the DW network annihilates immediately after the completion of the PT or evolves into the scaling regime.
\end{itemize}
For the values of the critical temperatures considered in \cite{Morgante:2022zvc}, the last two conditions are typically violated. Thus, even if the fluctuations manage to relax for the network to form, it will annihilate immediately. Assuming that all of the conditions above are satisfied, in the next section we estimate the GW signal from a DW network that reaches the scaling regime and then annihilates at later times before starting to dominate the energy budget of the universe. 

\section{Gravitational Waves from Annihilating Domain Walls}\label{Section: GW from annihilating domain walls}

Thus, provided that the conditions for having metastable DWs are satisfied, the annihilation temperature is given by equating the energy density of the DWs during scaling $\rho_{DW}\simeq \sigma_{DW} H(T_{ann})$ to the energy bias $\delta V$ to find 
\begin{equation}\label{annihilation DW}
    T_{ann} \simeq 2.27 \sqrt{\frac{M_{Pl,4d}T_{c}\epsilon_{\theta}}{N_{c}}}\,.
\end{equation}
DWs are dangerous cosmological relics since their energy density scales as $\rho_{DW}\sim a^{-1}$, which is slower than radiation scaling with $\rho_{rad}\sim a^{-4}$. Therefore, one needs to ascertain that the network annihilates before it starts to become the dominant energy component of the universe \cite{Zeldovich:1974uw}. The temperature at which the produced DWs would start to dominate the energy density of the universe is found by setting $\rho_{DW}= \rho_{rad}$ and one finds \begin{equation}
    T_{dom} \simeq 0.5 \sqrt{\frac{N_{c}\ T_{c}^{3}}{M_{Pl,4d}}}\,.
\end{equation} 
Hence, the condition which we have to impose is \begin{equation}
    T_{ann}>T_{dom}\,.
\end{equation}
 It should be noted that the percolation theory condition \eqref{percolation bound on theta} directly impacts the annihilation temperature, whereas the domination temperature is insensitive to $\theta_{UV}$. In fact, reducing $\epsilon_{\theta}$ allows us to delay annihilation for a fixed combination of $N_{c}$ and $T_{c}\,.$

As we want to reach scaling, we parametrize the critical temperature with a dimensionless parameter $x$, namely
\begin{equation}
    T_{c}= x \cdot \frac{\epsilon_\theta}{N_{c}}\, 2\cdot 10^{14}\text{ GeV}= x\cdot T_{c,m}(\theta_{UV},N_{c})\,.
\end{equation} 
Hence, to have a DW network which remains in the scaling regime for a sufficient time before annihilation, we require  $T_{ann}\ll T_{c}\,,$ leading to  $x\gg 10^{4}\,$. This provides us with a lower bound on the critical temperature, reinterpreted as a prefactor $x$ which needs to be sufficiently large. Interestingly enough, taking the extreme limit $T_{ann}\lesssim T_{c}$ one finds that the parameter $x$ is independent of $\theta_{UV}$ and $N_{c}\,.$ This implies that the separation between the critical temperature and the annihilation temperature can be made arbitrarily large by increasing $x$.  The upper limit on $x$ is set by requiring that we do not enter the DW domination era, i.e., $T_{ann}>T_{dom}\,.$ This leads to an upper bound on the critical temperature, i.e. on the parameter $x$. Thus, we obtain that the viable range of $x$ is 
\begin{equation}\label{range of interest}
    10^4\ll x\leq \frac{1.31 \times 10^{4}}{\sqrt{\epsilon_{\theta}}}\,. 
\end{equation}
Therefore, one needs to have sufficient tuning in $\theta$ (or $\epsilon_\theta$) both to have a sufficient separation of scales between the annihilation temperature and the critical temperature and to satisfy the percolation bound in \eqref{percolation bound on theta}. As an example, consider \eqref{percolation bound on theta} with $\xi=10$, then if we wish to have a separation of scales with $x\approx10^6$, we need to have $\epsilon_{\theta}\lesssim10^{-4}$.

As the DWs expand and eventually annihilate due to the volume pressure generated by the bias $\delta V$, they produce a direct source of anisotropic stress in the energy-momentum tensor, generating gravitational waves. The GWs are mainly emitted when the network annihilates, as that is when the total energy density $\rho_{DW}$ is the largest.  Our estimates bear some degree of uncertainty, as neither an explicit computation of the DW surface tension in IHQCD has been performed nor the explicit time to approach scaling. Hence, the estimates here should be thought of as an order-of-magnitude discussion. The magnitude and spectral shape of the GW spectrum from the annihilation of DWs are found from numerical simulations such as \cite{Notari:2025kqq,Babichev:2025stm} with its peak amplitude and frequency given by \begin{align}
\Omega_{\rm GW}^{\rm peak}h^{2} \simeq 2.8\times 10^{-18}\left(\frac{\sigma}{\text{TeV}^{3}}\right)^{2}\left(\frac{10\text{ MeV}}{T_{ann}}\right)^{4} ,\\
f_{p}\simeq 1.1 \times 10^{-9} \text{ Hz} \left(\frac{T_{ann}}{10 \text{ MeV}}\right).
\end{align}
For the spectral shape of the spectrum, we adopt the results from \cite{Ferreira:2022zzo,Notari:2025kqq} given as \begin{equation}
    \Omega_{\rm GW}h^2=\Omega_{\rm GW}^{\rm peak}h^{2}\times S\left(\frac{f}{f_{p}}\right)\,, \qquad S(x) = \frac{4}{x^{-3} + 3x}\,,
\end{equation} 
where the spectrum is governed by a causal $f^{3}$ IR tail and above the peak it scales as $f^{-1}$ found from simulations.

In our explicit model, the DW tension and annihilation temperature depend on $T_{c},N_{c}$ and $\epsilon_{\theta}=\pi-\theta_{UV}$. For the peak amplitude and peak frequency, parametric proportionality relations are found to be  
\begin{equation}\label{gw spectrum in terms of Tc}
\begin{split}
    \Omega_{\rm GW}^{\rm peak} h^{2}&\simeq 2.9\cdot 10^{-60} N_{c}^{4}\left( \frac{T_{c}}{\text{PeV}} \right)^{4}\frac{1}{\epsilon_\theta^{2}},
    \\ f_{p}&\simeq 1.23 \cdot 10^{5} \text{Hz}\left( \frac{T_{c}}{100 \text{TeV}} \right)^{1/2}\left(\frac{\epsilon_{\theta}}{N_{c}}\right)^{1/2}.
\end{split}
\end{equation}
In terms of the prefactor $x$ this can be written as
\begin{equation}\label{gw spectrum in terms of x}
\begin{split}
    \Omega_{\rm GW}^{\rm peak} h^{2}&\simeq 4.76\cdot 10^{-27} x^4 \epsilon_\theta^{2},
    \\ f_{p}&\simeq 54.71 \text{ GHz} \ \sqrt
x\left(\frac{\epsilon_\theta}{N_{c}}\right).
\end{split}
\end{equation}
From Eq.~\eqref{gw spectrum in terms of Tc}, it is easy to see that in order to maximize GW prospects one needs to have the DW annihilation close to the upper bound given by the temperature of DW domination. However, if one allows for a high value of $T_{c}$, the degree of tuning in $\theta_{UV}$ becomes greater in order to satisfy the upper bound on DW annihilation. Moreover, considering the upper bound in \eqref{range of interest}, from \eqref{gw spectrum in terms of x} we find that the peak amplitude is bounded by 
    \begin{equation}
        \Omega_{\rm GW}^{\rm peak}h^2\lesssim 1.4\cdot 10^{-10}.
    \end{equation}
Increasing $T_{c}$ also shifts the peak towards higher frequencies. However, this can be compensated for by making $N_c$ larger or getting $\epsilon_{\theta}$ closer to $0$, which would bring the signal to lower frequencies. In conclusion, to enter the range relevant for future GW experiments ~\cite{LIGOScientific:2016wof,Yagi:2011wg,Punturo:2010zz} 
\begin{equation}
\Omega_{\rm GW}^{\rm peak}\sim 10^{-13}, \quad  f_p\sim 10-100\text{ Hz},
\end{equation}
one would need to achieve significant fine-tuning between $\epsilon_{\theta}$, $N_c$, and $T_c$ (or $x$).\footnote{We should also emphasize that the value of $T_{c}$ needs to satisfy the bounds coming from reheating, regarding the maximal temperature of the universe $T_{max}\,.$ In single-field inflation with maximal reheating temperatures are given by $T_{max}\sim \mathcal{O}(3-5)\cdot 10^{15}\text{GeV}$~\cite{Planck:2018vyg}. Interestingly enough, in the regime where the GW amplitudes are the largest $T_{max}$ actually becomes a constraint of merit one needs to account for. For example, take again $\xi=10$, in the case when $x=10^{6}$ and $\epsilon_{\theta}=1.71\times10^{-4}$ (saturating $T_{ann}>T_{dom}$), to satisfy $T_{c}< T_{max}$\, one finds the bound $N_{c}\gtrsim 7$\,.}

Hence, for ideal phenomenological purposes, one would prefer to maximize the split between the critical temperature and the annihilation temperature to attempt to have two distinct peaks in the total GW spectrum, one from the PT and one from DWs annihilation. The best combination is to consider $\epsilon_{\theta} \ll1$, $N_{c}\gg 1$, and $T_{c}$ as high as possible within the bounds  to allow for the formation and adequate evolution of the DW network. Then, depending on how close $\epsilon_{\theta}$ is to $0$, it might be preferable to generate one large GW peak from DW annihilation, as the amplitude is maximized and the peak frequency is lowered for fixed combinations of $T_{c}$ and $ N_{c}$. If the initial critical temperature is sufficiently low, one can try to have both GW peaks in the ET/DECIGO regime for correlated signal searches, but this seems severely tuned within our setup.

\section{Discussion \& Conclusion}\label{Section: Discussion and Conclusions}

In this work, we showed that the inclusion of the $\theta$-angle reduces the range of supercooling available in the confining phase transition of $SU(N_c)$ Yang-Mills theories. When studying the effects of finite $\theta$ on the YM phase diagram, we pinpoint the lower limit where large $N_{c}$ approximations may be valid, and find the lower limit to be $N_{c,min} = 10\,$ where finite $N_{c}$ effects are directly relevant. This bound may serve as a general insight when having large $N_{c}$ arguments in mind, while working at moderately large but finite $N_{c}$. 

Additionally, since different bubbles may occupy different branches of Yang-Mills vacua, the DW network can form if the $\theta$-angle satisfies the bound from percolation theory. Provided the network forms, the bias is typically strong enough to annihilate the network immediately, effectively leaving no observational imprint in GWs unless significant fine-tuning is in place.
In the future, one can attempt to compute the explicit surface tension of the domain walls in our holographic construction. For example, this can be done by assuming a planar wall and the axion profile which depends on an additional spatial coordinate, i.e. $a(r,z)$, where $z$ is the coordinate perpendicular to the wall. However, as already discussed, we expect this to alter the surface tension by an $\mathcal{O}(1)$ factor, which appears to be irrelevant in our estimates.

Another effect that would be interesting to understand is how droplet formation would delay the initial completion of the PT and modify local DW formation, as surface tension would become spatially dependent. Our analysis suggests that DW formation in strongly coupled confinement transitions may be governed by a hierarchy of local equilibration and macroscopic damping scales. While local wall formation occurs rapidly, reheating and bulk-viscous effects may delay the emergence of a connected network. Quantifying these effects requires dedicated real-time simulations of planar walls in IHQCD with a $\theta$-angle.

A natural improvement in our calculations would be to include the full backreaction of the axion on the metric and the dilaton. We have identified the values of $N_c$ and $\theta_{UV}$ for which the backreaction should be included with the requirement that $T_c(\theta_{UV})>T_{min}$. Additionally, the backreaction becomes important far within the SBH branch, where it should make the free energy diagram smoothly approach zero. However, we do not expect our conclusions on the amount of supercooling to change significantly. The axion is different from the dilaton because it has a shift symmetry, i.e. $a\rightarrow a+\alpha$ for some constant $\alpha$, and the conserved quantity in question is the axion charge. Thus, the full inclusion of the axion backreaction may lead to the appearance of new saddles charged under this symmetry, such as Giddings-Strominger wormholes \cite{Giddings:1987cg} (see also e.g. \cite{Hebecker:2018ofv,Arkani-Hamed:2007cpn,Aguilar-Gutierrez:2023ril,Witten:2026twr,Jonas:2023ipa,Andriolo:2022rxc,Bergshoeff:2004pg,Hertog:2024nys,Marolf:2025evo,Loges:2022nuw,Jonas:2023qle,Held:2026huj}).  Whether axion wormholes are admitted in IHQCD and, if so, whether they contribute to the phase diagram of the dual gauge theory, remains, to our knowledge, an open question, which we leave for future work.

To summarize, although the cosmological imprint of the $\theta$-angle on confining phase transitions in $SU(N_c)$ Yang--Mills theories remains hidden in the absence of significant tuning, our analysis shows that finite-$\theta$ confinement transitions provide a concrete setting in which domain-wall production is governed by a hierarchy of dynamical processes. These
include vacuum assignment during bubble nucleation, local equilibration and reheating during bubble coalescence, and the macroscopic damping of thermal fluctuations. Thus, in strongly coupled non-conformal FOPTs, domain-wall production should not be viewed as an instantaneous consequence of percolation alone.

\section*{Acknowledgements}
The authors thank Pedro Schwaller for great discussions.
NR thanks Rishad Roshan for interesting discussions about DW production during FOPTs, Motoo Suzuki for valuable discussions on domain walls in non-abelian gauge theories, and Matthew McCullough for remarks about the $\theta$ angle on the lattice. BM acknowledges support from the INFN initiative APINE. NR  was supported in part by the European Union - NextGenerationEU through the PRIN Project “Charting unexplored avenues in Dark Matter” (20224JR28W).
EM acknowledges support from the Italian Ministry for University and Research (MUR) Rita Levi-Montalcini grant “New directions in axion cosmology”. EM and NR are supported by Istituto Nazionale di Fisica Nucleare (INFN) through the Theoretical Astroparticle Physics (TAsP) project.

\appendix
\section{Relevant timescales}\label{Relevant timescales}
In this Appendix, we elaborate a bit further for the reader interested in the local equilibration time scales during the onset of bubble coalescence.

\paragraph{Reaching $\tau_{\rm hydro}$}: Real-time simulations of planar shocks within the AdS/CFT framework in strongly coupled CFTs are mainly conducted in $\mathcal{N}=4$ SUSY YM theory \cite{Janik:2014kfa,Heller:2011ju,Heller:2012je}, where it was found that $\tau_{\rm hydro}\sim \mathcal{O}(1/T)$ considering Bjorken flow. This is the time scale at which, from an initial state, the system's temperature evolution is governed by viscous hydrodynamics. In particular, for $\tau\geq \tau_{\rm hydro}$ the diffusion length of the system is governed by the shear viscosity $\eta$, which is related to the entropy density as $\eta/s = \frac{1}{4\pi}$.  However, in non-conformal cases, $\tau_{\rm hydro}$ is found to be increased \cite{Attems:2016tby,Attems:2017zam} due to the presence of additional transport coefficients, primarily due to the presence of the bulk viscosity $\zeta$. Hence $\tau_{\rm hydro}^{\rm non CFT} \gtrsim (1.5-2) \tau_{\rm hydro}^{\rm CFT}$ roughly by the magnitude of the bulk viscosity $\zeta$ in relation to the shear viscosity. However, the studies reviewed in \cite{Berges:2020fwq} indicate that $\tau_{\rm hydro}^{\rm nonCFT}$ is still within the same order of magnitude as in the CFT case. As bulk and shear viscosity affect different channels of relaxation, this is not a straightforward claim. However, when calculating $\zeta/s$ using IHQCD as in \cite{Gursoy:2009kk} one finds $\zeta/s \sim \eta/s$ in the range $T_{c}>T>T_{min}$. This indicates that the sound attenuation constant which scales as
\begin{equation}
    \Gamma_{s} \simeq \frac{1}{sT_c}\left(\frac{4}{3}\eta + \zeta\right)=\frac{1}{T_{c}}\frac{7}{3\cdot 4\pi}= \frac{7}{12\pi T_{c}},
\end{equation} 
is increased by a factor of $\frac{7}{4}$ compared to its conformal value $\Gamma_{s}= \frac{1}{3\pi T_{c}}$, validating the qualitative statements above for the macroscopic hydrodynamization time scale of thermal fluctuations. 
    
For our purposes, the main implication is that the collision region quickly hydrodynamizes $\tau_{\rm hydro}\sim \mathcal{O}(1/T_{c})$, such that the subsequent evolution towards an effective LTE can be described by viscous hydrodynamics. When $\tau \sim \tau_{\rm eq}$ the system is described by an effective LTE regime containing ideal hydro plus first order viscous hydro whose transport coefficients are calculated within IHQCD.
To affirm our recent claims regarding the magnitude of the bulk viscosity $\zeta$ in relation to the shear viscosity $\eta$, we calculate the bulk viscosity coefficient $\zeta$ using the Eling-Oz formula \cite{Eling:2011ms}
\begin{equation}
    \frac{\zeta}{\eta} = \left(s \frac{d\phi_{h}}{ds} \right)^{2} = \frac{1}{9}A'(\phi_{h})^{-2} = \frac{1}{9}(\lambda_{h}A'(\lambda_{h}))^{-2}\,.
\end{equation}
This can be further simplified into an expression for the bulk viscosity to entropy ratio of the following form
\begin{equation}
    \frac{\zeta}{s} = \frac{\eta}{s}\left( \frac{1}{9\lambda_{h}^{2}A'(\lambda_{h})^{2}} \right) = \frac{1}{4\pi}\left( \frac{1}{3\lambda_{h}A'(\lambda_{h})} \right)^{2}, 
\end{equation}
In \cite{Buchel:2011wx} it was shown that the Eling-Oz formula in IHQCD matches well with the conventional approach of computing the IR limit of the two-point function of the trace of the stress energy tensor \cite{Gubser:2008sz,Gursoy:2009kk}. 
    
\begin{figure}[t!]
    \centering
    \includegraphics[width=1\linewidth]{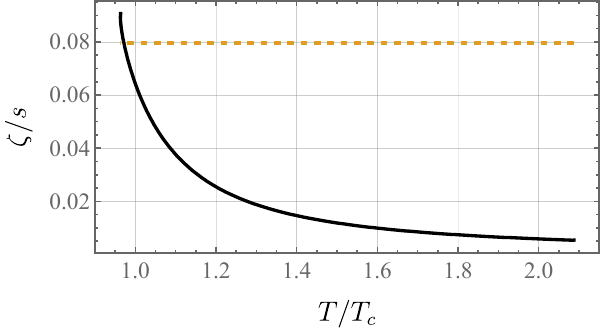}
    \caption{Bulk viscosity-to-entropy density ratio $\zeta/s$ in black compared to the fixed value for the shear viscosity-to-entropy density bound $\eta/s = 1/4\pi$ in yellow dashed.}
    \label{fig:entropy ratio}
\end{figure}

\paragraph{Reaching $\tau_{\rm eq}$:} We draw insights mainly from \cite{Attems:2016tby,Attems:2017zam}, which also showcases the timescale at which isotropization is achieved, in other words, when the different components of the pressure have become roughly equal. Interestingly enough, in \cite{Chesler:2016ceu,Attems:2016tby,Attems:2017zam,Attems:2018gou,Attems:2019yqn} it was found that isotropization is often achieved more slowly even though the average pressure may have relaxed. Furthermore, the fact that the viscous hydrodynamic description is applicable before isotropization is achieved is quite remarkable. 
    However, in our context of bubble collisions (assumed to be planar), this should instead translate into saying that the pressure gradient is becoming small.
    Nevertheless, for our purposes, it is more important to establish an estimate for when the system, after bubble coalescence, approaches an effective LTE regime, such that one can assign a local temperature and EoS sufficiently described by $T^{\mu\nu}(\bar{x}) \simeq (p+e)u^{\mu}u^{\nu} + \Pi^{\mu\nu}\,$ in which the viscous corrections $\frac{\eta}{p}<1\,, \frac{\zeta}{p}<1$. In the strongly coupled setting $\tau_{\rm eq}\sim \mathcal{O}(5-10)\tau_{\rm hydro}$ which, as we will see, is well within a Hubble time for our viable region of critical temperatures.
    
\paragraph{Reaching $\tau_{\rm micro}$:}  Microscopic thermalization of conformal strongly coupled QFTs was considered in \cite{Chesler:2011ds,Balasubramanian:2011ur}, where they looked at the evolution of the vacuum expectation value of the energy-momentum tensor and scalar correlation functions. From the gravity picture one studies the equilibration of a black hole geometry by studying its perturbed evolution and decay of the quasi-normal modes. In these scenarios, one finds that $\tau_{\rm micro}\sim \mathcal{O}(1/T)$, in particular, the thermalization of modes with high frequency and small momenta is fastest. In contrast, modes with both large momenta and frequency thermalize more slowly. In non-conformal scenarios considered in \cite{Buchel:2015saa,Janik:2015waa}, it was also found that the microscopic thermalization timescale still goes as $\tau_{\rm micro}^{\rm non CFT}\sim\mathcal{O}(1/T_c)$. This has been further corroborated in actual computations of QNMs for IHQCD in \cite{Alho:2020gwl}, which showcased that the thermalization timescale is of the order $\mathcal{O}(1/T_c) $    near the confinement PT. The dominant QNM is the one stemming from the lightest glueball mass, whose QNM is the slowest to equilibrate. The main result of \cite{Alho:2020gwl} was to show that in the PT regime it still decays on a timescale of $1/T_c\,.$ 

\bibliography{biblio}
\end{document}